\documentstyle [12pt]{article}

\parskip=\medskipamount                 
\def\ast{\displaystyle *}
\begin{document}

\bibliographystyle{unsrt}                                                    

\begin{center}
{\Large \bf BOUND STATES IN GALILEAN-INVARIANT\\ QUANTUM FIELD THEORY}
\footnote{Supported in part by the National Science Foundation;
email addresses, corley@umdhep.umd.edu, greenberg@umdhep.umd.edu}\\
[5mm]
S.R. Corley and O.W. Greenberg\\
{\it Center for Theoretical Physics\\
Department of Physics \\
University of Maryland\\
College Park, MD~~20742-4111}\\[5mm]
\end{center}

\vspace{2mm}

\begin{center}
{\bf Abstract}
\end{center}

We consider the nonrelativistic quantum mechanics of a model of two spinless
fermions interacting via a two-body potential.  We introduce quantum fields
associated with the two particles as well as the expansion of these fields in
asymptotic ``in'' and ``out'' fields, including such fields for bound states, in
principle.  We limit our explicit discussion to a two-body bound state.  In this
context we discuss the implications of the Galilean invariance of the model and,
in particular, show how to include bound states in a strictly 
Galilean-invariant quantum field theory.  

\newpage

{\bf I. Introduction}

The representations of the Galilean group relevant to quantum 
theory are
ray representations with a non-trivial phase discovered by Bargmann\cite{barg}.
The Bargmann phase leads to the Bargmann mass superselection rule that 
the mass of a
bound state must be exactly the sum of the masses of its 
constituents.  We analyze bound states in strictly Galilean-invariant theories 
taking 
account of the Bargmann phases.  Our technique is the 
Haag expansion\cite{haag} of the fields that appear in the Hamiltonian in 
normal-ordered 
products of asymptotic (in- or out-) fields.
We use the representation theory of the Galilean group due to Bargmann to
constrain the form of the Haag expansion and
 derive the Schr\"odinger equation for bound states, the unitarity relation
for elastic scattering and other relations in a unified way.  We derive the
relation between the in- and out- asymptotic fields for bound states by 
constructing the out-field as the asymptotic limit of the product of the fields
of the constituents at separated points integrated with the bound state 
amplitude that serves as the wavefunction.  

In Sec. II, we introduce the two-body model that we consider.  In Sec. III, we
derive the transformation properties of the Haag amplitudes under Galilean 
transformations.  We take care to show that
the Bargmann mass-dependent phases
that occur in Galilean-invariant theories
cancel so that we can consider breakup and rearrangement processes in which the
initial particles have different masses from the final particles.  As just
mentioned, the Bargmann
superselection rule requires that the sum of the masses that occur in the 
${\bf p}^2/2m$ kinetic terms is absolutely conserved in all processes.
In Sec. IV,  we investigate the 
anticommutation relations of the interacting fields in terms of their Haag 
expansions and obtain relations (as examples, the relation between the
bound-state amplitudes with different legs on- and off-shell, and 
elastic unitarity) among the amplitudes independent of the
specific dynamics of a given theory.
In Sec. V, we apply the NQA to the bound state problem and show 
that the Haag amplitude describing this state is just the 
Schr\"odinger wavefunction.  To our knowledge this is the first 
description of a
bound state in which Galilean invariance is strictly maintained.  
In Sec. VI, we apply the NQA to the two-body
scattering problem and show that the corresponding amplitude yields the 
T-matrix after removal of its off-shell leg.      
In Sec. VII, we construct the asymptotic fields for a
bound state as the integral with the bound-state wavefunction of the product of
the fields for the elementary constituents of the composite system.  Here we
differ from the proposals of Nishijima\cite{nish} and of Zimmermann\cite{zimm}
who construct bound states
as products of the constituent fields at the same point.
Section VIII concludes with a summary and the outlook for further research.

{\bf II. Two-Body Model}

We consider a model in the Heisenberg picture with two spinless 
nonrelativistic Fermi fields, $A({\bf x},t)$ and $B({\bf x},t)$, 
with the Hamiltonian
\begin{eqnarray}
H & = & \frac{1}{2m_{A}}\int\,d^{3}x\,
\nabla_{{\bf x}}A^{\dag}({\bf x},t)\cdot \nabla_{{\bf x}}A({\bf x},t)+
\frac{1}{2m_{B}}\int\,d^{3}x\,
\nabla_{{\bf x}}B^{\dag}({\bf x},t)\cdot \nabla_{{\bf x}}B({\bf x},t) 
\nonumber \\
  &   & \mbox{}
+\int\,d^{3}x\,d^{3}y\,
B^{\dag}({\bf y},t)A^{\dag}({\bf x},t)
V_{AB}(|{\bf x}-{\bf y}|)A({\bf x},t)B({\bf y},t);
\end{eqnarray}
for simplicity we assumed an $AB$ interaction, but no $AA$ or $BB$
interaction.  We assume $V$ is smooth, not too long ranged, and not too singular
at the origin.  We want $V$ to be well-behaved enough that the weak asymptotic
limit we introduce just below exists.  Since this is a very technical issue, we
are deliberately vague about the necessary conditions.  The literature on this
issue can be traced from articles and references in\cite{moller}.
The equation of motion for $A({\bf x},t)$ is
\begin{equation}
{\em i}\partial_{t}A({\bf x},t)
=-\frac{1}{2m_{A}}\nabla^{2}_{{\bf x}}A({\bf x},t)
+\int d^{3}y\,
B^{\dag}({\bf y},t)V_{AB}(|{\bf x}-{\bf y}|)B({\bf y},t)A({\bf x},t).\label{em}
\end{equation}
Some calculations are simpler in momentum space, therefore we define
\begin{equation}
A({\bf x},t) =\int\,d^{3}k dEe^{-{\em i}(Et-{\bf k}\cdot {\bf x}})
\tilde{A}({\bf k},E),
\end{equation}
\begin{equation}
V_{AB}(|{\bf x}-{\bf y}|) 
= \frac{1}{(2\pi)^3}\int\,d^{3}q\,e^{{\em i}{\bf q}\cdot 
({\bf x}-{\bf y})}\tilde{V}_{AB}(|{\bf q}|).
\end{equation}
Transforming the equation of motion to momentum space yields
\[ (E_A-\frac{{\bf k}_A^2}{2m_{A}})\tilde{A}({\bf k}_A,E_A)=
\int\,dE_Bd^{3}k_BdE^{\prime}_Bd^{3}k^{\prime}_BdE^{\prime}_Ad^{3}k^{\prime}_A
 \nonumber \] 
\[ \times \delta(E_A+E_{B}-E^{\prime}_{B}-E^{\prime}_{A})
\delta({\bf k}_A+{\bf k}_B-
{\bf k}^{\prime}_A-{\bf k}^{\prime}_B) \nonumber \]
\begin{equation}
\times \tilde{B}^{\dag}({\bf k}_B,E_B)\tilde{V}_{AB}(|{\bf k^{\prime}}_B
-{\bf k}_B|)
\tilde{B}({\bf k}^{\prime}_B,E^{\prime}_B)
\tilde{A}({\bf k}^{\prime}_A,E^{\prime}_A).
\end{equation}
The asymptotic (in- or out-) fields for (possibly composite) particles are
characterized by their rest energy $E$, 
mass $m$, and spin $J$.  We will suppress the spin in what follows.  
The
definitions of the asymptotic fields associated with the interacting field
$A({\bf x}, t)$ are
\begin{equation}
A^{in~(out)}({\bf x}, t)=lim_{t^{\prime} \rightarrow \mp \infty}\int 
{\cal D}({\bf x}-{\bf y}, t-t^{\prime};0,m_A)A({\bf y},t^{\prime})d^3y,
                                                                \label{as}
\end{equation}
where the limit is the weak limit of the smeared operators, and 
\begin{equation}
{\cal D}({\bf x},t;E,m)=\frac{1}{(2 \pi)^{3}}\int d\omega d^3k
\delta(\omega-E-\frac{{\bf k}^2}{2m}) e^{-i\omega t+i{\bf k}\cdot {\bf x}}.  
                                                                \label{cal}
\end{equation}
The asymptotic fields for $B({\bf x}, t)$ are defined in an analogous way.  
We give the
definition of the asymptotic fields for composite particles in Sec. VII.  The
asymptotic limit in momentum space is often useful in calculations,
\begin{equation}
\tilde{A}^{in~(out)}({\bf k},E)=lim_{t^{\prime} \rightarrow \mp \infty}
\delta(E-{\bf k}^2/2m) \int dE^{\prime} e^{i(E-E^{\prime})t^{\prime}} 
\tilde{A}({\bf k}, E^{\prime}).                                  \label{asmom}
\end{equation}
Either form of the definition of the asymptotic limits makes clear that the
asymptotic fields obey the free equations of motion,
\begin{equation}
i\partial_t C^{in(out)}({\bf x}, t) 
= (E-\frac{1}{2m}\nabla^{2})C^{in(out)}({\bf x}, t) 
\end{equation}
and also the free field  anticommutation or
commutation relations,
\begin{equation}
[C^{in(out)}({\bf x},t),C^{\dag in(out)}({\bf y},t^{\prime})]_{\pm} 
= {\cal D}({\bf x}-{\bf y},t-t^{\prime};E,m).           
\end{equation}
Note that
\begin{equation}
{\cal D}({\bf x},0;E,m)=\delta({\bf x}),~~ \forall E, m.
\end{equation}

Using translation invariance, the Haag expansion of the interacting field 
$A({\bf x},t)$ in terms of in-fields takes the
following form in position space (with an analogous expansion for the $B$ 
field) 
\begin{equation}
A({\bf x},t)=A^{in}({\bf x},t) + \sum_{i} \int d^{3}x_{B} d^{3}x_{i} 
f_{B;i}({\bf x}-{\bf x}_B,t-t_B;{\bf x}-{\bf x}_i,t-t_i) 
\nonumber \\
\end{equation}
\[ \times B^{\dag \, in} ({\bf x}_{B},t_B)(ABi)^{in}({\bf x}_{i},t_i) \]
\[ + \int d^3 x_B d^3 x^{\prime} d^3x^{\prime}_B 
f_{B;AB}({\bf x}-{\bf x}_B,t-t_B;{\bf x}-{\bf x}^{\prime},
t-t^{\prime}_B;
{\bf x}-{\bf x}^{\prime}_B,t-t^{\prime}_B) \]
\begin{equation}
 \times B^{\dag \, in} ({\bf x}_B,t_B) 
A^{in} (x^{\prime},t^{\prime}) B^{in}({\bf x}^{\prime}_B) +\cdots. 
                                                               \label{he} 
\end{equation}
Because the asymptotic fields obey free equations, the Haag amplitudes obey 
free equations in each individual argument.  
A simple way to see this is to note
that the convolution in position space becomes a product in momentum space, so
that the momentum arguments of the Haag amplitudes are multiplied by energy
shell delta functions contained in the asymptotic fields.  Thus only the
on-energy shell part of the Haag amplitudes enters and the position space Haag
amplitudes obey the free equations.  Since both the asymptotic fields and the
Haag amplitudes obey the free equations, 
the integrals are independent of the times $t_B,t_i$ and of
$t_B,t^{\prime},t^{\prime}_B$ because of the translation
invariance of the Schr\"odinger scalar products.
We label the Haag amplitude that is
the coefficient of a product of (asymptotic) creation and annihilation operators
by the labels of the operators; the two-body 
$(AB)$ bound state in
level $i$ is labeled by $i$.

We define 
\begin{equation}
C^{in}({\bf x},t)=(2 \pi)^{-3/2}\int dE d^3k \delta(E-E_C-{\bf k}^2/2m_C)
e^{-iEt+i{\bf k}\cdot {\bf x}}c^{in}({\bf k}),
\end{equation}
\begin{equation}
[c^{in}({\bf k}),c^{\dagger~in}({\bf l})]_+=\delta({\bf k}-{\bf l})  \label{ft}
\end{equation}
\begin{equation}
f_{B;i}({\bf x},t; {\bf x}^{\prime},t^{\prime})= \]
\[ \frac{1}{(2\pi)^3}\int d^3k_1 d^3k_2 
\exp(i\frac{{\bf k}_1^2}{2m_B}t-i{\bf k}_1\cdot{\bf x}
-i(-\epsilon_i+\frac{{\bf k}_2^2}{2m_{AB}})t^{\prime}+i{\bf k}_2\cdot {\bf
x}^{\prime})
\tilde{f}_{B;i}({\bf k}_1,{\bf k}_2)
\end{equation}
and similar definitions for other Fourier transforms chosen so 
that powers of 
$2\pi$ are absent from the momentum-space formulas.  The result is
\[ \tilde{A}({\bf k},E)=(2 \pi)^{-3/2}a^{in}({\bf k})
\delta(E-\frac{{\bf k}^2}{2m_A}) \]
\[+ \int \,d^{3}k_{B}d^{3}k_{i}\,
\delta(E+\frac{{\bf k}^2_{B}}{2m_B}+\epsilon_i-
\frac{{\bf k}^2_{i}}{2m_{AB}})\delta({\bf k}+{\bf k}_{B}-{\bf k}_{i})
\tilde{f}_{B;i}({\bf k}_B;{\bf k}_{i})  \]
\[ \times a^{in \dag}({\bf k}_B)a^{in}_i({\bf k}_{i}) \]
\[ + \int \,d^{3}k_{B}d^{3}k^{\prime}_{B}d^{3}k^{\prime}\,
\delta(E+\frac{{\bf k}_{B}^2}{2m_B}-\frac{{\bf k}^{\prime~2}}{2m_A}
-\frac{{\bf k}^{\prime ~2}_{B}}{2m_B})\delta({\bf k}+{\bf k}_{B}
-{\bf k}^{\prime}-{\bf k}^{\prime}_{B}) \]
\begin{equation}
\times \tilde{f}_{B;AB}({\bf k}_B;{\bf k}^{\prime},{\bf k}^{\prime}_B)
a^{in \dag}({\bf k}_{B})a^{in}({\bf k}^{\prime})a^{in}({\bf k}_{B})
+\cdots.
\end{equation}
(We use the abbreviation $m_{AB}=m_A+m_B$.)
Note that we are expanding in terms of in-fields;
there are analogous expansions in terms
of out-fields.  In the next section we derive the constraints on the $f$'s that
follow from Galilean invariance.

{\bf III. Galilean Invariance}

Bargmann showed that the unitary projective representations 
(i.e., representations up to a factor) of the 
Galilean group that occur in the quantum mechanics of nonrelativistic particles
cannot be reduced to vector (i.e., true) representations.  This contrasts with
the situation for the Poincar\'e and Lorentz groups,
and--indeed--most other physically interesting groups, where the representations
can be reduced to true representations.  As already mentioned twice, the 
explicit mass
parameter in the phases leads to the Bargmann superselection rule that the 
sum of the masses (that appear in the kinetic terms) must be conserved in every
process.  Nonetheless, bound states can be formed and particles can be created
and annihilated, provided the Bargmann superselection rule is obeyed.  

Note, for example, that if we were to assign rest energies $m_A$ and $m_B$ to
particles $A$ and $B$ then a bound state of these particles with binding energy 
$\epsilon$ would have energy $E=m_{AB}-\epsilon+
{\bf k}^2/2 m_{AB}$, rather than $E=m_{AB}-\epsilon+
{\bf k}^2/2 (m_{AB}-\epsilon)$ as one might expect from the
nonrelativistic limit of a relativistic bound state with rest energy 
$m_{AB}-\epsilon$.  
Another manifestation of this effect is that for this 
same bound state the momentum would transform under Galilean boosts as
${\bf k} \rightarrow {\bf k}+m_{AB}{\bf v}$, rather than as
${\bf k} \rightarrow {\bf k}+(m_{AB}-\epsilon){\bf v}$.

If the 
projective representation has the form
\begin{equation}
U(G_2)U(G_1)=\omega(G_2,G_1)U(G_2G_1)
\end{equation}
then another projective representation is equivalent to this if the other
representation has the factor system $\omega^{\prime}(G_2,
G_1)=[\phi(G_2)\phi(G_1)/\phi(G_2G_1)]\omega(G_2,G_1)$, 
where $\phi$ has modulus
one.  This arbitrariness allows simplification of some formulas.

Bargmann gives as the Galilean transformation of a 
nonrelativistic scalar wave function,
\begin{equation}
(T(G)\psi)({\bf x},t)=e^{-i\theta(G,({\bf x},t))}\psi(G^{-1}({\bf x},t)),
\end{equation}
where $G({\bf x},t)=(R{\bf x}+{\bf v}t+{\bf a}, t+b)$ and 
$\theta(G,({\bf x},t))=m(\frac{1}{2}{\bf v}^2t-{\bf v}\cdot {\bf x})$.   
The Galilean transformation is labeled by $({\bf a}, b, R,{\bf v})$, 
where ${\bf a}$ and
$b$ are space and time translations, $R$ is a rotation and ${\bf v}$ is a
boost.
To infer the corresponding transformation for a nonrelativistic scalar
field, we require
\begin{equation}
U(G)A(\psi)U^{\dagger}(G)=A(\psi_G),~~\psi_G({\bf x},t)=(T(G)\psi)({\bf x},t)
=e^{-i\theta(G,({\bf x},t))}\psi(G^{-1}({\bf x},t)),
\end{equation}
\begin{equation}
A(\psi)=\int A({\bf x},t)\psi({\bf x},t)dt d^3x.
\end{equation}
We find that
\begin{eqnarray}
U(G)A({\bf x},t)U^{\dagger}(G)&=&e^{-i\theta_A(G,G({\bf x},t))}
A(G({\bf x},t)), \nonumber \\
\theta_A(G,G({\bf x},t))&=&m_A
[\frac{1}{2}{\bf v}^2(t+b)-{\bf v}\cdot (R{\bf x}+{\bf v}t+{\bf a})].\label{f}
\end{eqnarray}
If the field has spin $s$, then $A$ on the left hand side is replaced by $A_i$
and $A$ on the right hand side is replaced by $\sum_j A_j D^{(s)}_{ji}(G)$,
where $D^{(s)}$ is a representation of $SU(2)$, which is the little group in
this case.
The corresponding transformation holds for $B$ with $m_B$ replacing $m_A$.
Asymptotic fields transform the same way.    The 
implications of the transformation law for the Haag amplitudes 
is found by transforming the interacting field in two ways: (1) act on the 
Haag expansion with $U(G)$ as in the left
hand side of Eq.(22) and redefine the integration variables, and (2) multiply
the Haag expansion by the phase factor on the right hand side of Eq.(22) and
replace $({\bf x},t)$ by $G({\bf x},t)$.  
The two amplitudes $f_{B;i}$ and $f_{B;AB}$ obey
\[f_{B;i}(G({\bf x}_A-{\bf x}_B,t_A-t_B);
G({\bf x}_A-{\bf x}_i,t_A-t_i))=\]
\begin{equation}
 e^{i\theta_A(G,G({\bf x}_A,t_A))+i\theta_B(G,G({\bf x}_B,t_B))
-i\theta_{AB}(G,G({\bf x}_i,t_i))}
f_{B;i}({\bf x}_A-{\bf x}_B,t_A-t_B);{\bf x}_A-{\bf x}_i,t_A-t_i),
\end{equation}
\[f_{B;AB}(G({\bf x}_A-{\bf x}_B,,t_A-t_B));G({\bf x}_A-{\bf x}^{\prime}_A,t_A-
t^{\prime}_A),G({\bf x}_A-{\bf x}^{\prime}_B,t_A-t^{\prime}_B)=\]
\[ e^{i\theta_A(G,G({\bf x}_A,t_A))+i\theta_B(G,G({\bf x}_B,t_B))
-i\theta_{A}(G,G({\bf x}^{\prime}_A,t^{\prime}_A))-
i\theta_B(G,G({\bf x}^{\prime}_B),
t^{\prime}_B))} \]
\begin{equation}
\times f_{B;AB}({\bf x}_A-{\bf x}_B,t_A-t_B;x_A-{\bf x}^{\prime}_A,t_A-
t^{\prime}_A),{\bf x}_A-{\bf x}^{\prime}_B,t_A-t^{\prime}_B).
\end{equation}
Note that $\theta_{AB}$ is independent of the bound state $i$ because of the 
Bargmann mass
superselection rule.  The combination of phases in the first of these is
\[\theta_A(G,G({\bf x}_A,t_A))+\theta_B(G,G({\bf x}_B,t_B))-
\theta_{AB}(G,G({\bf x}_i,t_i))=\]
\begin{equation}
-\frac{1}{2}{\bf v}^2(m_At_A+m_Bt_B-m_{AB}t_i)
-{\bf v}\cdot R(m_A{\bf x}_A+m_B{\bf x}_B-m_{AB}{\bf x}_i).
\end{equation}
The transformation law is {\it not} satisfied by having a delta function in
the space and time coordinates identifying the coordinates $({\bf x}_i,t_i))$ 
with the
center-of-mass of particles $A$ and $B$, although at equal times such a delta
function does occur for the space coordinates.  The way in which the
transformation laws are satisfied is best seen in momentum space, where
the corresponding transformations in momentum space are 
\begin{equation}
(V(G)\phi)({\bf k},E)=
e^{-i\Omega(G,({\bf k},E))}\phi(G^{-1}({\bf k},E)),
\end{equation}
\begin{equation}
\Omega(G, ({\bf k},E))=({\bf k}-m{\bf v})\cdot{\bf a}-
(E-\frac{1}{2}m{\bf v}^2)b,
\end{equation}
where $G({\bf k},E)=(R{\bf k}+m{\bf v}, E+{\bf v}\cdot R{\bf k}+
\frac{1}{2} m {\bf v}^2)$, and 
$G^{-1}({\bf k},E)=(R^{-1}({\bf k}-m{\bf v}), 
E-{\bf k}\cdot {\bf v}+\frac{1}{2}m{\bf v}^2)$. 
The momentum space transformation law for the field is induced in parallel with
the derivation of the position space law.  The result is
\begin{equation}
W(G)A({\bf k},E)W^{\dagger}(G)=e^{-i\Omega_A(G,-G({\bf k},E))}
A(G({\bf k},E), \label{15}
\end{equation}
where $\Omega_A(G,-G({\bf k},E))=(E+{\bf v}
\cdot R{\bf k})b-R{\bf k}\cdot {\bf a}$.
In the transformation law for the Haag amplitudes, all the phase factors 
cancel and the result for--say--the second term in
the Haag expansion is what one would expect naively,
\begin{equation}
\tilde{f}_{B;i}({\bf k}_B;{\bf k}_i)=
\tilde{f}_{B;i}(R({\bf k}_B-m_B{\bf v});R({\bf k}_i-m_{AB}{\bf v})).
\end{equation}
Thus we can choose the ${\bf v}={\bf k}_i/m_{AB}$ so that the bound-state 
momentum vanishes and eliminate the second argument of $f_{B;i}$,
\begin{equation}
\tilde{f}_{B;i}({\bf k}_B;{\bf k}_{i})=
\tilde{f}_{B;i}({\bf k}_B-\frac{m_B}{m_{AB}}{\bf k}_i,{\bf 0})
\equiv \tilde{F}_{B;i}({\bf k}_B-\frac{m_B}{m_{AB}}{\bf k}_i). \label{gi}
\end{equation}
For the spinless case, 
$\tilde{F}_{B;i}({\bf k})=\tilde{F}_{B;i}(R{\bf k}).$
All these results are exact, valid in any Galilean frame.  The 
extension to fields
with spin is straightforward.  It is worth noting that the Poincar\'e
transformation law in a relativistic theory is simpler than the Galilean
transformation law we have just derived for a nonrelativistic theory, 
because the Bargmann phase is absent for the Poincar\'e group.   

Taking account of Galilean invariance, the position-space Haag amplitude is
\[f_{B;i} ({\bf x},t;{\bf x}^{\prime},t^{\prime}) = 
(2\pi)^{-3} \int d^{3}k  d^{3}k_{i} 
exp[i (m_{B} + \frac{1}{2m_{B}}({\bf k} +
\frac{m_{B}}{m_{AB}}{\bf k}_{i})^{2}) t - i ({\bf k} +
\frac{m_{B}}{m_{AB}}   {\bf k}_{i}) \cdot {\bf x}] \]
\begin{equation}
\times exp[-i (- \epsilon_i + \frac{{\bf k}^2_{i}}{2m_{AB}})
t^{\prime} + i {\bf k}_{i} \cdot {\bf x}^{\prime}]
\times \tilde{f}_{B;i} ({\bf k};{\bf 0}). 
\end{equation}
The integral over ${\bf k}_i$ can be done, but the result is complicated and
not useful, except when all times are equal, in which case the result is both
simple and useful,
\begin{equation}
f_{B;i}({\bf x}_A-{\bf x}_B;{\bf x}_A-{\bf x}_i)=
\delta({\bf x}_{i}-\frac{m_A {\bf x}_A+m_B {\bf x}_B}{m_{AB}})
F_{B;i}({\bf x}_A-{\bf x}_B),               \label{29}
\end{equation}
\begin{equation}
F_{B;i}({\bf x})=\int d^3k e^{-i {\bf k}\cdot {\bf x}} 
\tilde{f}_{B;i}({\bf k};{\bf 0}).
                                         \label{et}
\end{equation}
Using the constraints due to Galilean invariance, the Haag expansion in
$x$-space at equal times takes the form

\[ A({\bf x})=A^{in}({\bf x})+\sum_i\int F_{B;i}({\bf x}-{\bf x}_B)
B^{\dagger in}({\bf x}_B)(ABi)^{in}
(\frac{m_A{\bf x}+m_B{\bf x}_B}{m_{AB}})d^3x_B \]
\[ +\int d^3r^{\prime} d^3r F_{B;AB} ({\bf r^{\prime}};{\bf r})
B^{\dagger in}({\bf x}- {\bf r}^{\prime}) A^{in}
({\bf x} + \frac {m_B ({\bf r} - {\bf r}^{\prime})}{m_{AB}})
B^{in} ({\bf x}- \frac{m_A {\bf r} + m_B{\bf r}^{\prime}}{m_{AB}}) \]
\begin{equation}
+\cdots.
\end{equation}
In momentum space, the expansion is
\[ \tilde{A}({\bf k}, E)=\frac{1}{(2 \pi)^{3/2}}
a^{in}({\bf k})\delta(E-\frac{{\bf k}^2}{2m_A}) \]
\[ +\int d^3k_B
\delta(E+\frac{{\bf k}_B^2}{2m_B}+\epsilon_i-
\frac{({\bf k}+{\bf k}_B)^2}{2m_{AB}})
\tilde{F}_{B;i}(\frac{m_A{\bf k}_B
-m_B{\bf k}}{m_{AB}})  b^{in\dagger}({\bf k}_B)
c^{in}_i({\bf k}+{\bf k}_B)_i \]
\[+\int d^3k_B d^3p_B d^3p 
\delta(E+\frac{{\bf k}_B^2}{2m_B}-
\frac{{\bf p}^2}{2m_A}-\frac{{\bf p}_B^2}{2m_B})\delta({\bf k}+{\bf k}_B
-{\bf p}-{\bf p}_B)\]
\begin{equation}
\times \tilde{F}_{B;AB}(\frac{m_A{\bf k}_B-m_B{\bf k}}{m_{AB}};
\frac{m_A{\bf p}_B-m_B{\bf p}}{m_{AB}})b^{in~ \dagger}({\bf k}_B)a^{in~
\dagger}(p)b^{in}({\bf p}_B) +\cdots .
\end{equation}
Here $c^{in}_i$ is the annihilation operator for the bound state of $A$ and $B$
in state $i$.

{\bf IV. Two-Body Bound State}

To derive the equation for the two-body bound state, we insert the Haag
expansion Eq.(\ref{he}) in the equation of motion Eq.(\ref{em}), 
renormal order, and equate the
coefficients of the terms with the operators 
$B^{\dagger in}(AB_i)^{in}$.  After
commuting or anticommuting with the relevant in-fields, the result is
\begin{equation}
(i\frac{\partial}{\partial t}+\frac{1}{2m_A}{\bf \nabla}^2_{x}
-V(|{\bf x}-{\bf x}_B|))f_{B;i}({\bf x}-{\bf x}_B,t-t_B;
{\bf x}-{\bf x}_i,t-t_i)=0.
\end{equation}
It is convenient to eliminate the time derivative by using 
$\partial/\partial t=-\partial/\partial t_B
-\partial/\partial t_i$, the independence of the 
Schr\"odinger scalar product on the time and the free equations satisfied by the
in-fields 
to find free equations for the $t_B$ and $t_i$ dependences of $f_{B;i}$.
The results are
\begin{equation}
(i\frac{\partial}{\partial t_B}+\frac{1}{2m_B}{\bf \nabla}^2_{x_B})
f_{B;i}=0,
\end{equation}
\begin{equation}
(i\frac{\partial}{\partial t_i}-\epsilon_i-
\frac{1}{2m_{AB}}{\bf \nabla}^2_{x_i})f_{B;i}=0.
\end{equation}
The equation without time derivatives is
\begin{equation}
[-\frac{1}{2m_A}{\bf \nabla}^2_{x}-\frac{1}{2m_B}{\bf \nabla}^2_{x_B}+
V(|{\bf x}-{\bf x}_B|)]f_{B;i}=(\epsilon_i
-\frac{1}{2m_{AB}}{\bf \nabla}^2_{x_i})f_{B;i}.
\end{equation}
Now using Eq.(\ref{29}) we find the usual Schr\"odinger equation for 
$F_{B;i}$,
\begin{equation}
[-\frac{1}{2\mu}{\bf \nabla}^2_{r_{AB}}+
V({\bf r}_{AB})]F_{B;i}=-\epsilon_iF_{B;i},~~\frac{1}{\mu}=\frac{1}{m_A}+
\frac{1}{m_B}, 
\end{equation}
where the reduced mass enters.
This establishes that $F_{B;i}$ is the Schr\"odinger wave function of the bound
state.
Note that the bound-state amplitude is given
{\it exactly} in any reference frame in terms of the amplitude in the rest frame
of the bound state.  (The corresponding statement also holds for other 
amplitudes, as well as for relativistic theories.)  

{\bf V.  Two-Body Scattering}

Two-body scattering is described in position space at equal times by the 
amplitude 
\begin{equation}
f_{B;AB} ({\bf x}_A - {\bf x}_B,0; 
{\bf x}_A - {\bf y}_A,0, {\bf x}_B - {\bf y}_B,0) = F_{B;AB}
({\bf x}_A - {\bf x}_{B}; {\bf y}_A - {\bf y}_B) \delta
({\bf R^{\prime}} - {\bf R}),
\end{equation}
\begin{equation}
F_{B;AB} ({\bf x}; {\bf y}) = (2 \pi)^{-3/2}
\int d^3k^{\prime}d^3k \tilde{f}_{B;AB}({\bf k}^{\prime}; -{\bf k},
{\bf k}) exp~i[-{\bf k}^{\prime}\cdot ({\bf x}_A - {\bf x}_B) +
{\bf k}\cdot({\bf y}_A - {\bf y}_B)],
\end{equation}
$${\bf R}^{\prime} = \frac{m_A {\bf x}_A + m_B{\bf x}_B}
{m_{AB}},~~ \hspace{.2in} {\bf R} =
\frac{m_A{\bf y}_A + m_B{\bf y}_B}{m_{AB}}.$$
We prefer to discuss two-body scattering in momentum space, using
the amplitude 
$\tilde{f}_{B;AB}({\bf k}_{B};{\bf p}_{A},{\bf p}_{B})$ which is the
coefficient of the term 
$b^{in \dag}_{B}({\bf k}_B)a^{in}({\bf p}_{A})b^{in}({\bf p}_{B})$ 
in the Haag expansion of 
$A({\bf k},E)$.  The procedure for finding the equation for 
$\tilde{f}_{B;AB}$ is analogous 
to that for the two-body bound state amplitude.  We find
\begin{eqnarray}
(\frac{{\bf p}_{A}^{2}-({\bf p}_A+{\bf p}_B-{\bf k}_{B})^{2}}{2m_{A}}
+\frac{{\bf p}_{B}^{2}-{\bf k}_{B}^{2}}{2m_{B}})
\tilde{f}_{B;AB}({\bf k}_{B};{\bf p}_{A},{\bf p}_{B}) &=&\nonumber \\
\tilde{V}_{AB}(|{\bf k}_{B}-{\bf p}_{B}|)
+\int\,d^{3}k_B^{\prime}\,\tilde{V}_{AB}(|{\bf k}_B-{\bf k}_B^{\prime}|)
\tilde{f}_{B;AB}({\bf k}_B^{\prime};{\bf p}_{A},{\bf p}_B).
\end{eqnarray}
Galilean invariance relates $\tilde{f}_{B;AB}$ at arbitrary 
momenta to itself in the center-of-mass, 
\begin{equation}
\tilde{f}_{B;AB}({\bf k}_{B};{\bf p}_{A},{\bf p}_{B}) = 
\tilde{f}_{B;AB}(R({\bf k}_{B}-m_B{\bf v});R({\bf p}_{A}-m_A{\bf v}),
R({\bf p}_{B}-m_B{\bf v})).
\end{equation}
By choosing ${\bf v}=({\bf p}_A+{\bf p}_B)/m_{AB}$,
we can replace $\tilde{f}_{B;AB}$ by a function of one fewer variable,
\begin{equation}
\tilde{f}_{B;AB}({\bf k}_{B};{\bf p}_{A},{\bf p}_{B})=  
\tilde{F}_{B;AB}({\bf k};{\bf p}),   \label{42}
\end{equation} 
where here and below,
${\bf k}=(m_A{\bf k}_B-m_B{\bf k}_A)/m_{AB}$,
${\bf p}=(m_A{\bf p}_B-m_B{\bf p}_A)/m_{AB}$ and we used conservation of
momentum to introduce ${\bf k}_A$.  The momenta ${\bf p}$
and ${\bf k}$ are the 
center-of-mass momenta of particle $B$ in the initial and
the final state, respectively.
The elastic scattering equation becomes
\begin{equation}
\frac{1}{2\mu}({\bf p}^2-{\bf k}^2)
\tilde{F}_{B;AB}({\bf k};{\bf p})
=\tilde{V}(|{\bf k}-{\bf p}|)+\int d^3k^{\prime}
\tilde{V}({\bf k}-{\bf k}^{\prime})\tilde{F}_{B;AB}
({\bf k}^{\prime};{\bf p}),
\end{equation}
The solution is the Born series,
\begin{equation}
\tilde{F}_{B;AB}({\bf k};{\bf p})=
\tilde{G}_R({\bf k};{\bf p})\tilde{V}(|{\bf k}-{\bf p}|)+ \nonumber\\
\end{equation}
\begin{equation}
\tilde{G}_R({\bf k};{\bf p})\int d^3k^{\prime}
\tilde{V}(|{\bf k}-{\bf k^{\prime}}|)\tilde{G}_R({\bf k^{\prime}};{\bf p})
\tilde{V}(|{\bf k^{\prime}}-{\bf p}|)+ \cdots,
\end{equation}
$\tilde{G}_R({\bf k};{\bf p})=
[({\bf p}^2-{\bf k}^2)/2 \mu-i\epsilon]^{-1}$.

The amplitude $\tilde{F}_{B;AB}$ 
is closely related to the $T$-matrix element for $AB$
scattering. The $S$-matrix element is
\begin{equation}
S({\bf k}_{A}, {\bf k}_{B};{\bf p}_{A},{\bf p}_{B}) =
\langle 0|b^{out}({\bf k}_{B})a^{out}({\bf k}_{A})
a^{in~\dag}({\bf p}_{A})b^{in~\dagger}({\bf p}_{B})|0\rangle.
\end{equation}
In order to eliminate the out-fields in terms of the in-fields we need 
the definitions, given above in Eq.(6),
\begin{equation}
A^{in(out)}({\bf x},t)=\lim_{\tau \rightarrow \mp \infty}\int_{t^{\prime}
=\tau}\,d^{3}x^{\prime}\,{\cal D}({\bf x}-{\bf x}^{\prime},t-t^{\prime};
m_{A},m_{A})A({\bf x}^{\prime},t^{\prime}),
\end{equation}
where ${\cal D}$ was defined in Eq.(\ref{cal}).
The nonrelativistic analog of the reduction formula follows from calculating
$\int d^3x^{\prime}dt^{\prime} \partial/ \partial_{t^{\prime}}
{\cal D}({\bf x}-{\bf x}^{\prime},t-t^{\prime};m_A,m_A)A({\bf x}^{\prime},
t^{\prime})$ in two ways: 
performing the integral and carrying out the derivative.  The result
\cite{moller} is
\begin{equation}
A^{out}({\bf x},t)-A^{in}({\bf x},t)=\int\,d^{3}x^{\prime}dt^{\prime}
\,{\cal D}({\bf x}-{\bf x}^{\prime},t-t^{\prime};m_{A},m_A)
(\partial_{t^{\prime}}-
\frac{i}{2m_{A}}\nabla^{2}_{{\bf x}^{\prime}})A({\bf x}^{\prime},t^{\prime}).
\end{equation}
Fourier transforming this yields
\begin{equation}
\frac{1}{(2 \pi)^{3/2}}(a^{out}({\bf k})-a^{in}({\bf k}))=-2 \pi i
(E-\frac{{\bf k}^{2}}{2m_{A}})A({\bf k},E).  \label{in-out}
\end{equation}
Note that a factor of $\delta(E-{\bf k}^{2}/2m_{A})$ has been removed
from this equation; thus the right-hand-side is non-vanishing (and there is
scattering) only when $A({\bf k},E)$ has a pole at 
$E-{\bf k}^{2}/2m_{A}=0$.  Since 
$a^{\dagger~out}({\bf k})|0\rangle=a^{\dagger~in}({\bf k})|0\rangle$ 
for stable particles,
the only out operator in the $S$-matrix element
$\langle 0| b^{out}({\bf k}_B)a^{out}({\bf k}_A)A^{\dagger in}({\bf p}_A)
b^{\dagger in}({\bf p}_B)|0 
\rangle$ that must be eliminated using Eq.(\ref{in-out}) is
$a^{out}({\bf k}_{A})$.  The result is
\[ S({\bf k}_A, {\bf k}_B;{\bf p}_A, {\bf p}_B) = \delta({\bf k}_A-{\bf p}_A)
\delta({\bf k}_B-{\bf p}_B) 
-2 \pi i
\delta(\frac{{\bf k}_A^2}{2m_A}
+\frac{{\bf k}^2_B}{2m_B}-\frac{{\bf p}^2_A}{2m_A}
-\frac{{\bf p}^2_B}{2m_B}) \]
\begin{equation}
\times \delta({\bf k}_A+{\bf k}_B-{\bf p}_A-{\bf p}_B)
(\frac{{\bf k}_A^2}{2m_A}+\frac{{\bf k}^2_B}{2m_B}-\frac{{\bf p}^2_A}{2m_A}
-\frac{{\bf p}^2_B}{2m_B})\tilde{F}_{B;AB}({\bf k};{\bf p}),
\end{equation}
where again ${\bf k}$ and ${\bf p}$ are defined below Eq.(\ref{42}).  
Thus the reduced $T$-matrix 
 for elastic scattering on the momentum shell\cite{gold-watson} is 
\begin{equation}
t({\bf k}_A,{\bf k}_B;{\bf p}_A,{\bf p}_B)=
(\frac{{\bf p}_A^2}{2m_A}
+\frac{{\bf p}_B^2}{2m_B}-\frac{{\bf k}_A^2}{2m_A}-
\frac{{\bf k}_B^2}{2m_B})
\tilde{F}_{B;AB}({\bf k};{\bf p}).
\end{equation}
We emphasize that because the Haag amplitude is the scattering amplitude with
one leg off shell, it contains the information necessary for calculations in 
the three-body sector. This contrasts with the on-shell scattering amplitude,
which does not suffice for such calculations.

{\bf VI. Anticommutation Relations}

In this section we 
show that the canonical (equal time) 
anticommutation relations of the Lagrangian fields imply general 
relations among Haag
amplitudes, independent of the equations of motion of the specific theory.
For example, the vanishing of the canonical anticommutator $[A,B]_+$ at equal 
times, considered for the coefficient of the bound state in-field for state $i$,
gives
\begin{equation}
F_{A;i}({\bf y}-{\bf x})=F_{B;i}({\bf x}-{\bf y})\equiv F_i({\bf x}-{\bf y}) 
\end{equation}
where we took $(ABi)^{in}({\bf R})=-(BAi)^{in}({\bf R})$ because of the Fermi
statistics of $A$ and $B$.  This shows that the 
apparent asymmetry in the treatment of 
the
constituents of the bound state, due to the fact that the Haag amplitude that 
serves as the two-body wave
function of the $(AB)$ bound state in the Haag expansion of the $A$ field has 
the $A$ particle off-shell and the $B$
particle on-shell, while these roles are interchanged for the amplitude for the 
same bound state in the Haag expansion of the $B$ field, is not a real
asymmetry.  These two amplitudes determine each other uniquely.  The analogous
result for the off-shell elastic scattering amplitudes is
\begin{equation}
F_{B;AB} ({\bf x}-{\bf y};{\bf r})=F_{A;BA} ({\bf y}-{\bf x};-{\bf r}) \equiv
F_{AB} ({\bf x}-{\bf y};{\bf r})
.
\end{equation}
Again the two apparently different off-shell amplitudes uniquely determine each
other.

The consequence for elastic scattering is
\begin{eqnarray}
\lefteqn{t({\bf k}_A,{\bf k}_B;{\bf p}_A,{\bf p}_B)-
(t({\bf p}_A,{\bf p}_B;{\bf k}_A,{\bf k}_B))^{\ast}=}  \nonumber \\
& & (2 \pi)^{5/2} \int d^3q_A d^3q_B \delta(\frac{{\bf k}_A^2}{2m_A}
+\frac{{\bf k}^2_B}{2m_B}-\frac{{\bf q}^2_A}{2m_A}
-\frac{{\bf q}_B^2}{2m_B})\delta({\bf k}_A+{\bf k}_B-{\bf q}_A-{\bf q}_B) 
\nonumber  \\
& & \times t({\bf k}_A,{\bf k}_B;{\bf q}_A,{\bf q}_B)
(t({\bf p}_A,{\bf p}_B;{\bf q}_A,{\bf q}_B))^{\ast},
\label{elun}
\end{eqnarray}
where ${\bf k}$ and ${\bf p}$ are as defined below Eq.(\ref{42}).  This 
is elastic unitarity.

The
canonical anticommutator $[A,A^{\dagger}]_+$ at equal times leads to a 
generalization of unitarity,
\[ \frac{1}{(2 \pi)^{3/2}}
(\tilde{F}_{B;AB}({\bf k};{\bf p})+\tilde{F}_{B;AB}^{\ast}({\bf p};{\bf k}))\]
\begin{equation}
=\sum_i\tilde{F}_{B;i}({\bf k})\tilde{F}^{\ast}_{B;i}({\bf p}) + 
\int d^3q\tilde{F}_{B;AB}({\bf k};{\bf q})
\tilde{F}^{\ast}_{B;AB}({\bf p};{\bf q}),
\end{equation}
where again ${\bf k}$ and ${\bf p}$ are as defined below Eq.(\ref{42}) and we
have used momentum conservation, ${\bf k}_A+{\bf k}_B={\bf p}_A+{\bf p}_B$.  
By taking the
appropriate limit, we recover the elastic unitarity relation, Eq.(\ref{elun}).
Taking into account the relations between the Haag amplitudes in the expansions
of $A$ and of $B$, these are all the independent 
two-body relations obtained from the 
anticommutation relations.

There are also quadratic relations between the amplitudes for the
$(ABi)$ and $(ABj)$ bound states and the amplitudes for the breakup of these
bound states due to scattering with the $A$ or $B$ particle.  
Since this involves 
a higher sector, we do not give this relation here.

{\bf VII. Construction of the asymptotic field for the bound state}

In this section we show how to construct the asymptotic field for the bound
state from a product of Lagrangian fields.  
The procedure is to multiply the appropriate Lagrangian fields at separated 
space points, integrate 
with the bound-state amplitude in the relative coordinate,
and take the asymptotic limit.
The result is
\[(ABi)^{in~(out)}({\bf x},t)=\lim_{\tau \rightarrow \mp\infty}
\int_{t^{\prime}=\tau}\,
d^{3}x^{\prime}\,{\cal D}({\bf x}-{\bf x}^{\prime},t-t^{\prime};
-\epsilon_i,m_{AB})F^{\ast}_{i}({\bf w})\]
\begin{equation}
\times \frac{1}{2} [B({\bf y}-\frac{m_A}{m_{AB}}{\bf w},t^{\prime}),
A({\bf y}+\frac{m_B}{m_{AB}}{\bf w},t^{\prime})]_-d^3w.
\end{equation}
A straightforward calculation shows these limits are 
$(ABi)^{in~(out)}({\bf x},t)$ 
for $\tau
\rightarrow \mp \infty$ and the leading terms for $\tau
\rightarrow \pm \infty$ are $(ABi)^{out~(in)}({\bf x},t)$.  
This is what we expect.  The
higher terms in the Haag expansion for 
$(ABi)^{out~(in)}(({\bf x},t)$ 
are in a higher sector that we don't discuss here.

{\bf VIII. Summary and outlook for further work}

We have derived many results of the nonrelativistic quantum mechanics of
two-particle systems in a unified way with particular attention to Galilean
invariance, 
taking into account the fact that the representations of the Galilean
group in quantum mechanics are necessarily representations up to a factor,
rather than vector representations.  We established the physical interpretation
of the Haag amplitudes:  the Haag amplitude for the simplest term with the
two-body bound-state operator is precisely the Schr\"odinger wave function of
the two-body bound state.  This interpretation will carry over to explicitly
covariant relativistic theories, 
where the corresponding Haag amplitude will be a
three-dimensional object, but will be covariant.  Of course in the relativistic
case, a bound state that is mainly a two-body state will also have amplitudes to
be composed of more particles.  We constructed the asymptotic field for 
a composite particle as the weak limit
of a product of the fields of the constituent particles weighted with the 
bound-state amplitude of the composite particle.  
We plan later to apply the N quantum formalism described here
to several-particle systems,
including scattering processes involving bound states and rearrangement
collisions.  In these cases, this formalism differs markedly from the usual
methods, such as the Faddeev analysis of three-body problems.  We do not
discuss here the problems that arise when the number of particles increases 
without bound; for example, in
the thermodynamic limit.  See Narnhofer and Thirring\cite{nt} for a discussion.
 The use of our 
techniques in relativistic theories has been considered in\cite{grs} among other
places.  This reference shows that, at least in the weak-coupling approximation,
the technique we discuss here can be used to find bound states in relativistic
theories.  We have also solved the Nambu--Jona-Lasinio model in one-loop
approximation with the Haag expansion\cite{njl}.  Although calculations based on
the operator field equations are not the most popular way to study relativistic
theories, the references just cited show that this can be a useful way to study
such theories.  
We are presently studying approximations in which we don't assume
weak coupling in collaboration with M. Malheiro and Y. Umino.  We hope that 
this method will serve as an alternative to the
Bethe-Salpeter equation in relativistic problems.
We also plan to construct variational
principles based on the Haag expansion for both nonrelativistic and relativistic
theories.

{\bf Acknowledgement}

We thank Eli Hawkins for pointing out that the precise relation
of unitarity follows from requiring $[A^{out},A^{\dagger~out}]_+=
[A^{in},A^{\dagger~in}]_+$.

\end{document}